\documentclass[traditabstract]{aa}
\usepackage{txfonts}
\usepackage{graphicx}
\usepackage{subfig}
\usepackage{natbib}
\usepackage{subfig}
\bibpunct{(}{)}{;}{a}{}{,}

\usepackage[colorlinks=true, citecolor=blue, linkcolor=blue, breaklinks=true]{hyperref}
\newcommand{\hi } {{\rm H}\,{\small\rm I} \,}
\newcommand{\hii } {{\rm H}\,{\small\rm II} \,}
\newcommand{\hiV} {{\rm H}\,{\small\rm I}}
\newcommand{\kms} {km~s$^{-1}$ \,}
\begin{document}

\title{Dynamics of Starbursting Dwarf Galaxies: I~Zw~18.}
\author{Federico Lelli\inst{1}
\and Marc Verheijen\inst{1}
\and Filippo Fraternali\inst{1,}\inst{2}
\and Renzo Sancisi\inst{1,}\inst{3}}

\institute{Kapteyn Astronomical Institute, University of Groningen, Postbus 800, 9700 AV, Groningen, The Netherlands \\
\email{lelli@astro.rug.nl}
\and Department of Astronomy, University of Bologna, via Ranzani 1, 40127, Bologna, Italy
\and INAF - Astronomical Observatory of Bologna, via Ranzani 1, 40127, Bologna, Italy}

\date{}

\abstract{
I~Zw~18 is a prototype Blue Compact Dwarf (BCD), characterized by a strong starburst
and extremely low metallicity ($Z \sim 0.02$ $Z_{\odot}$). It has long been
considered a candidate young galaxy in the Local Universe, but recent studies indicate
the presence of old stars.

We analysed archival VLA observations of the 21 cm line and found that the \hi associated
to the starburst region forms a compact fast-rotating disk. The \hi column densities
are very high, up to $\sim$50-100~M$_{\odot}$~pc$^{-2}$ ($\sim0.6-1.2\times10^{22}$ atoms~cm$^{-2}$).
The rotation curve is flat with a steep rise in the inner parts, indicating the presence
of a strong central concentration of mass. Mass models with a dark matter halo show that
baryons may dominate the gravitational potential in the inner regions. A radial
inflow/outflow motion of $\sim$15~km~s$^{-1}$ is also present.

I~Zw~18 appears structurally different from typical dwarf irregulars in terms of gas
distribution, stellar distribution and dynamics. It may be considered as a ``miniature''
high-surface-brightness disk galaxy. These dynamical properties must be tightly related
to the starburst. They also shed new light on the question of the descendants of BCDs.

There is also extended \hi emission towards the outlying stellar complex
I~Zw~18~C and a $\sim$13.5~kpc \hi tail. An interaction/merger
between gas-rich dwarfs is the most likely explanation for the starburst.}

\titlerunning{Dynamics of Starbursting Dwarf Galaxies: I~Zw~18.}
\authorrunning{Lelli et al.}

\maketitle

\section{Introduction}

Blue compact dwarfs (BCDs) are low-mass galaxies that are experiencing a starburst. They are
usually characterized by small physical sizes ($\sim$ 2-3 kpc), low metallicities ($0.2 \lesssim
Z/Z_{\odot} \lesssim 0.02$) and relatively large amounts of gas (M$_{\rm{HI}}$/L$_{\rm{B}}\gtrsim$ 1).
The question was raised whether they are young galaxies undergoing their first burst of
star formation \citep{SearleSargent1972}, but several studies based on surface brightness
and color profiles (e.g. \citealt{GilDePaz2005}) as well as color-magnitude
diagrams of resolved stellar populations (e.g. \citealt{Tosi2009}) have demonstrated that
BCDs contain also old stars, with ages $>$2-3~Gyr. The star-formation histories of the nearby BCDs,
as derived using color-magnitude diagrams (e.g. \citealt{Tosi2009}), show that the starburst
is a short-lived phenomenon, typically sustained for $\sim$100~Myr (e.g. \citealt{McQuinn2010}).
Thus, BCDs are \textit{transition-type dwarfs}, but it is not clear whether there are
evolutionary connections with dwarf Irregulars (dIrrs), Spheroidals (dSphs) and/or Ellipticals
(dEs) (e.g. \citealt{Papaderos1996, vanZee2001}). Also, the mechanisms that trigger, sustain
and quench the starburst activity are not understood.

Regarding the \hi distribution and kinematics, various studies (e.g. \citealt{vanZee1998b, vanZee2001})
highlighted two striking properties of BCDs: i) they show strong concentrations of \hi to the starburst
region near the galaxy centre; ii) they usually have steep central velocity gradients. Both properties
are \textit{not} observed in more quiescent dIrrs (e.g. \citealt{Swaters2002}). This suggests a close
connection between the starburst, the compact distribution of baryons (gas and stars) and the \hi kinematics.
The nature of the steep velocity gradients is not clear and two main interpretations are possible:
i) fast rotation (e.g. \citealt{vanZee2001}), ii) gaseous inflows/outflows (e.g. \citealt{Kobulnicky2008},
\citealt{Cannon2004}). Fast rotation in the inner regions would indicate the presence of a strong
concentration of mass, that may be either luminous or dark. Gaseous inflows/outflows would be
linked to fuelling processes and/or feedback mechanisms.

On larger scales, BCDs usually show extended and diffuse \hi structures, that may form reservoirs
for fueling the starburst. Generally, two different kinds of structures are observed:
i) extended \hi disks in regular rotation, e.g. NGC~2915 \citep{Elson2010}, NGC~2366 \citep{Oh2008};
ii) complex filamentary structures, e.g. II~Zw~40 \citep{vanZee1998b}, NGC~5253 \citep{Kobulnicky2008}.
The study of these extended \hi structures can provide key information on the triggering mechanism
(external vs internal processes), the properties of the progenitor galaxies (gas-rich dIrrs vs
gas-poor dEs/dSphs), and the possible presence of massive gas inflows/outflows.

\begin{table*}[thbp]
\caption{VLA observing parameters}
\label{tab:obs}
\centering
\begin{tabular}{l l l l l}
\hline
\hline
Project Code & Array Conf. & Observing Dates                              & Time on Source & Calibrators                 \\
\hline
AC0710       & A           & 8, 9, 14, 18, 19, 28 Nov. \& 4, 6 Dec. 2004  & 33.3 h         & 0834+555, 1331+305, 0542+498 \\
AZ0074       & B           & 26 Oct. \& 3, 15 Dec. 1995                   & 15.5 h         & 0834+555, 1331+305, 0542+498, 0137+331 \\
AP264        & C           & 27 Aug. 1993                                 & 6.6  h         & 0834+555, 1331+305 \\
AP264        & D           & 22 Dec. 1993                                 & 2.3  h         & 0834+555, 1331+305 \\
\hline
\end{tabular}
\end{table*}
\begin{table*}[thbp]
\caption{\hi Datacubes}
\label{tab:cubes}
\centering
\begin{tabular}{l c c c c c c c c}
\hline
\hline
Cube     & Robust & UV Taper    & Pixel size        & Synthesized beam    & Beam P.A. & Smoothed beam         & $\Delta V$  & Rms noise\\
     &   & (k$\lambda$)    & (arcsec$\times$arcsec) & (arcsec$\times$arcsec) & (degrees) & (arcsec$\times$arcsec) & (km s$^{-1}$) & (mJy/beam)\\
\hline
Low Res. & 0 & 10          & $3\times3$       & $14.6\times14.4$   & 32.7         & $20 \times 20$ & 5.2 & 0.40 \\
Intermediate Res. & 0 & 60 & $0.5\times0.5$   & $3.3\times2.9$     & 89.3         & $5 \times 5$   & 5.2 & 0.26 \\
High Res. & -1 & 150       & $0.25\times0.25$ & $1.5\times 1.4$    & -76.3        & $2 \times 2$   & 10.4 & 0.16 \\
\hline
\end{tabular}
\end{table*}

We present here a \hi study of I~Zw~18, the BCD prototype (e.g. \citealt{Zwicky1966, SearleSargent1972}).
I~Zw~18 is one of the most metal-poor galaxies known (12+log(O/H)$\sim$7.2, \citealt{Izotov1999})
and has long been considered a candidate young galaxy in the Local Universe, formed within the
last 0.5 Gyr (e.g. \citealt{Papaderos2002, Izotov2004}). However, \citet{Aloisi2007}, using
HST observations, detected stars older than 1-2~Gyr and ruled out the possibility that I Zw 18
is a truly primordial galaxy (as also suggested by e.g. \citealt{Aloisi1999}, \citealt{Ostlin2005}).
The intense star-formation activity started only $\sim$20 Myr ago (e.g. \citealt{Aloisi1999}).
Two key questions remain open: What triggered the starburst? Why is the metallicity so low?

Previous \hi studies \citep{Viallefond1987, vanZee1998} showed that I~Zw~18 is characterized
by a strong central concentration of \hi and a steep velocity gradient, as is typical for BCDs.
Also, the optical galaxy is surrounded by an extended \hi envelope that was described by
\citet{vanZee1998} as ``a fragmenting \hi cloud in the early stages of galaxy evolution''.

We analysed archival \hi data in order to: i) investigate the relation between the gas
distribution and the starburst activity; ii) clarify the nature of the velocity gradient;
iii) study the structure, kinematics and origin of the extended gas.

\section{Data reduction \& analysis \label{sec:reduction}}

We analysed public \hi data taken from the VLA archive. The observations were carried out between
1993 and 2004, using the VLA in all 4 configurations (see table \ref{tab:obs}). Data from
the B, C, and D configurations were presented by \citet{vanZee1998}. In this new analysis, we included
also data taken in 2004 with the high-resolution A-array configuration. The correlator was used
in 2AD mode, with a total bandwidth of 0.8 MHz ($\sim$165 km s$^{-1}$). An on-line Hanning taper
was applied to the data, producing 127 spectral line channels with a width of 6.3 kHz ($\sim$1.3 km s$^{-1}$).

The raw UV data were interactively flagged, calibrated and combined using the AIPS package and following
standard VLA procedures. Next, the UV data were mapped using a robust weighting technique \citep{Briggs1995} 
and various Gaussian baseline tapers to attenuate the longest baselines. We built three datacubes
with different spatial resolutions and pixel sizes by using different combinations of the Robust
parameter and the taper FWHM (see table \ref{tab:cubes}). After various trials, we chose the
combinations that minimize sidelobes and wings in the beam profiles.

After the Fourier transform, the datacubes were analysed using the Groningen Imaging Processing SYstem (GIPSY)
\citep{vanderHulst1992}. Continuum maps were constructed by averaging line-free channels. Because of the
small bandwidth of the observations, few line-free channels were available and the resulting
continuum-subtracted datacubes showed correlated noise in the spectral direction. Thus, we
constructed a continuum map by using a mask, defining the area of \hi emission in every
channel and averaging, for each spatial pixel, all the channels without \hi signal. The masks were constructed 
by smoothing the datacubes both in velocity (by a factor 4) and spatially (by a factor $\sim$3, i.e.
at 45$''$, 10$''$ and 5$''$ for the low, intermediate and high resolution data, respectively)
and clipping at 2.5$\sigma_{s}$ (where $\sigma_{s}$ is the noise in the smoothed cubes).
The masks were inspected channel by channel and remaining noise peaks were blotted out.

The use of a mask for the continuum subtraction may have the disadvantage that the noise is no longer uniform
across the channel maps, as a different number of channels is used at every pixel to build the continuum map.
Thus, we built signal-to-noise (S/N) maps for every channel (similarly to \citealt{Verheijen2001}) and
calculated a pseudo-1$\sigma$ contour by averaging the values of the pixels with 0.75$<$S/N$<$1.25.
The resulting pseudo-1$\sigma$ level is close to that obtained by calculating the noise in a box without signal,
suggesting that the noise is still almost uniform.

The channel maps were cleaned \citep{Hogbom1974} down to 0.5$\sigma$, using the masks to define the search areas
for the clean-components, which were then restored with a Gaussian beam of the same FWHM as the antenna pattern.
Next, to boost the S/N-ratio, the cubes were smoothed in velocity to a resolution of 5.2 km~s$^{-1}$ 
(10.4 km~s$^{-1}$ for the high-resolution data) and spatially to 20$''$, 5$''$ and 2$''$ for the low,
intermediate and high resolution data, respectively. Table \ref{tab:cubes} summarizes the properties
of the cubes.

Total \hi maps were constructed by summing the signal inside the clean-masks. A pseudo-3$\sigma$ contour
was calculated following \citet{Verheijen2001}. Velocity fields were built by fitting a Gaussian function
to the \hi line profiles. Fitted Gaussians with a peak intensity less than 2.5$\sigma$ and a FWHM smaller
than 5.2 \kms were discarded; remaining noise in the velocity fields (i.e. signal outside the pseudo-3$\sigma$
contour of the total \hi maps) was blotted out. The \hi line profiles are quite broad and asymmetric,
thus the velocity fields must be considered just as a rough indication of the global kinematics.
Our kinematical analysis is based on Position-Velocity diagrams (Sect. \ref{sec:cubes})
and on 3D models of the observations (Sect. \ref{sec:kin}).

\begin{figure*}[t!]
\centering
\subfloat{\includegraphics[width=18.3cm]{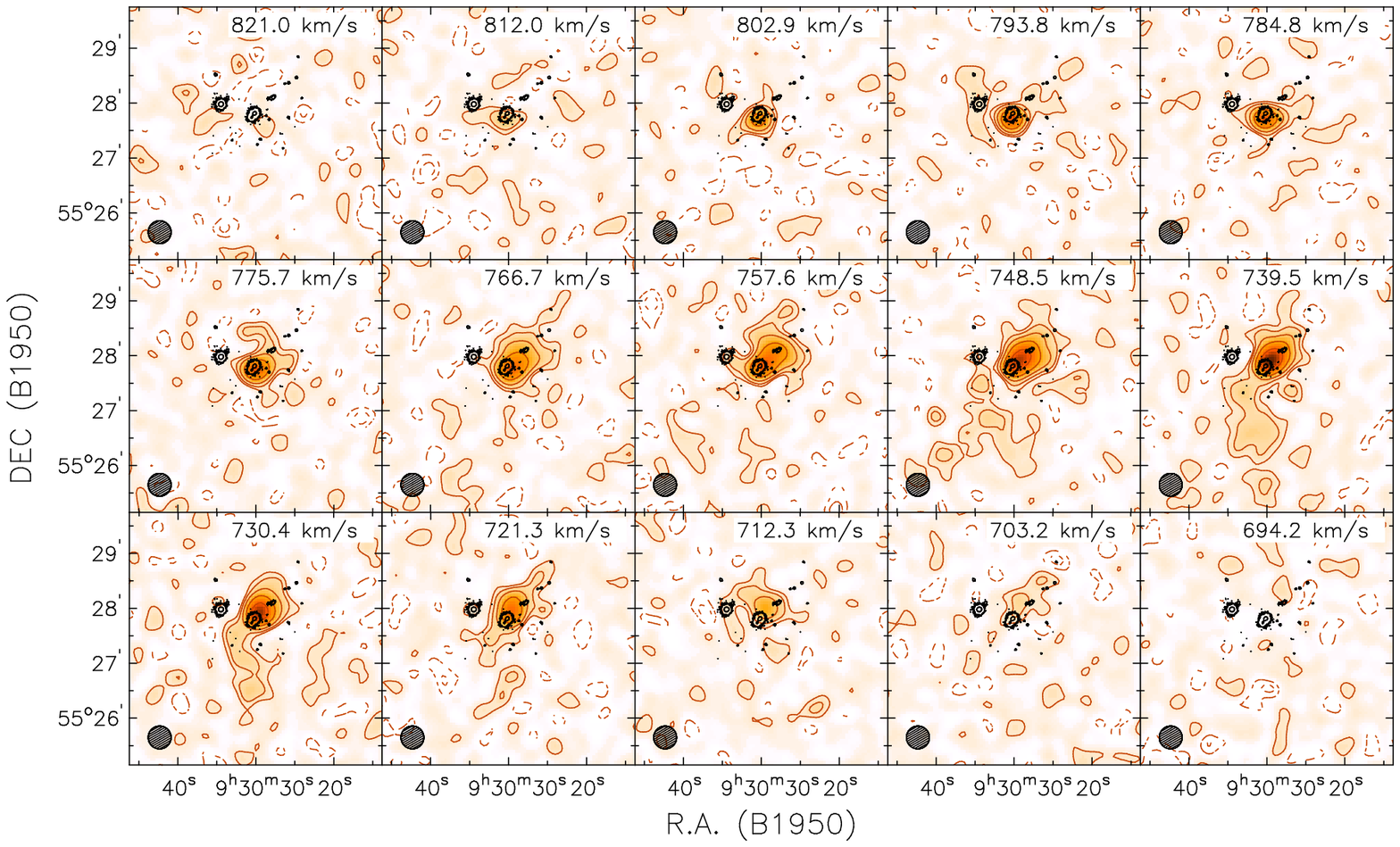}}\\
\subfloat{\includegraphics[width=18.3cm]{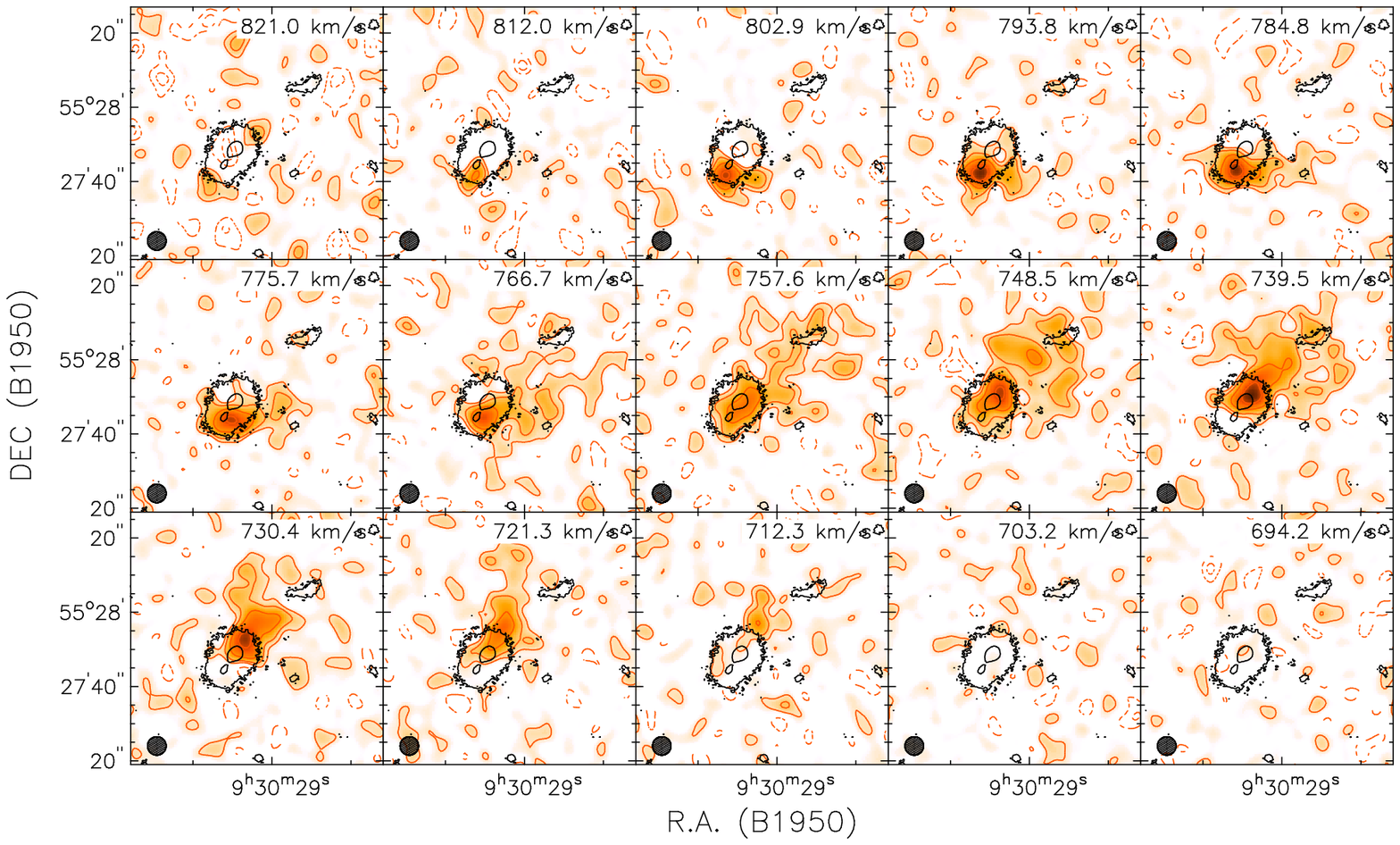}}
\caption{Channel maps at a resolution of 20$''$ (top) and 5$''$ (bottom).
Red-solid contours are at 1.5, 3, 6, 12, 24 $\times$ $\sigma$.
Red-dashed contours are at -3, -1.5 $\times$ $\sigma$.
Black contours show two isophotes of a B-band image (from \citealt{GilDePaz2003});
the object to the North-East (R.A.~=~9h~30m~34s, DEC~=~55$^{\circ}$~28$''$)
is a foreground star.}
\label{fig:s20}
\end{figure*}

\section{\hi distribution and kinematics \label{sec:cubes}}

\begin{figure*}[thp]
\centering
\includegraphics[width=16cm]{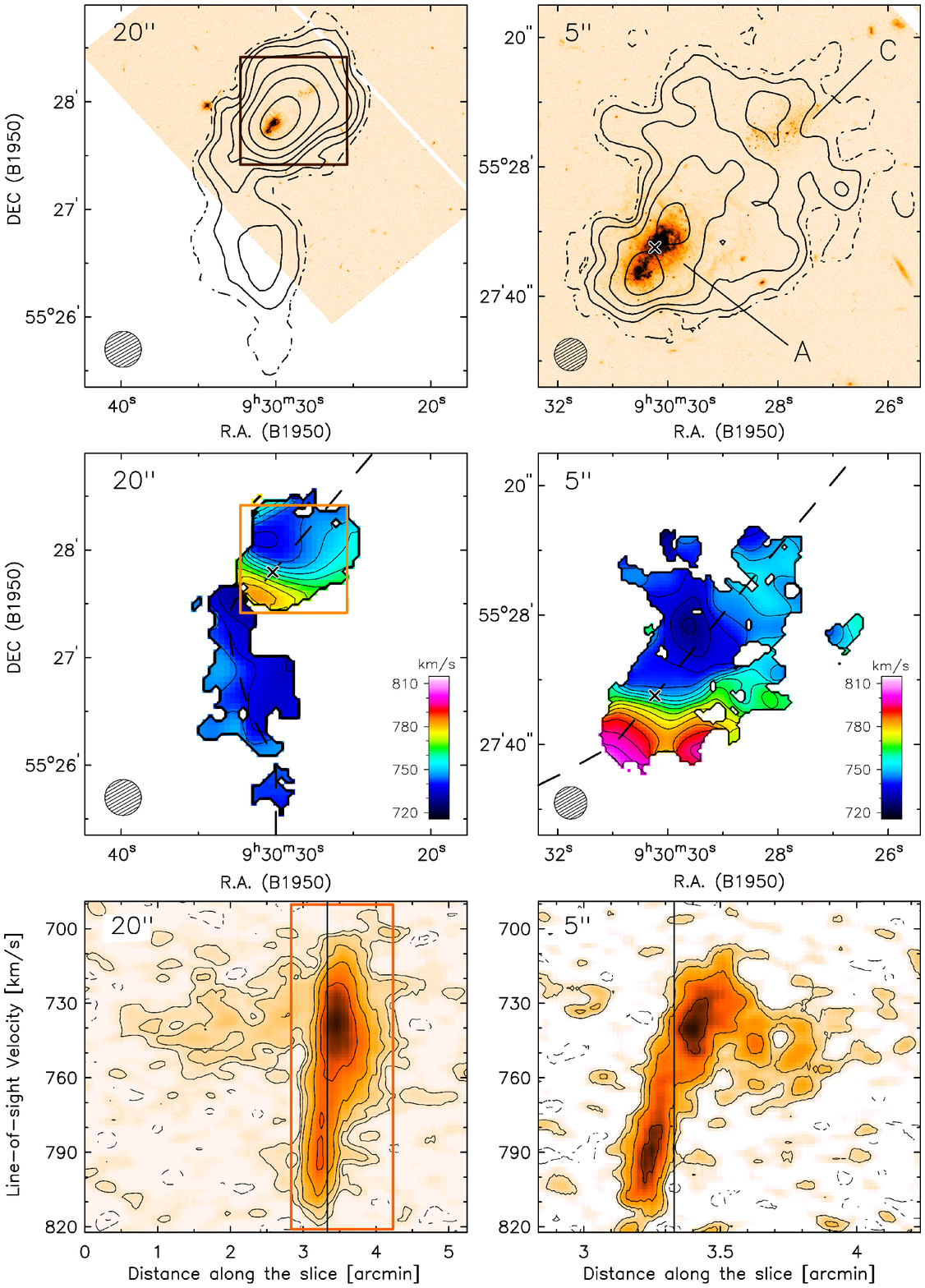}
\caption{
\textbf{Top}: integrated \hi maps at a resolution of 20$''$ (left) and 5$''$ (right), overlayed
on a HST image (from \citealt{Aloisi2007}). The box in the left panel shows the area covered
by the right panel. The dashed line corresponds to the pseudo-1.5$\sigma$ density contour.
In the map at 20$''$, contours are at 0.25 (dashed), 0.5, 1, 2, 4, 8, 16 $\times$
10$^{20}$ atoms cm$^{-2}$. In the map at $5''$, contours are at 3 (dashed), 6, 12,
24, 48 $\times$ 10$^{20}$ atoms cm$^{-2}$. The circle shows the beam size.
\textbf{Middle}: velocity fields at a resolution of 20$''$ (left) and 5$''$ (right). The box in
the left panel shows the area covered by the right panel. Contours range from 722.4 to 805.6
km~s$^{-1}$, with steps of 5.2~km~s$^{-1}$. The circle shows the beam size. The dashed
line shows the path followed to obtain the position-velocity diagram.
\textbf{Bottom}: position-velocity diagrams at a resolution of 20$''$ (left) and 5$''$ (right).
Contours are at -1.5 (dashed), 1.5, 3, 6, 12, 24 $\times$ $\sigma$. The box in the left panel
shows the region covered by the right panel. The vertical line corresponds to the
cross in the velocity fields.}
\label{fig:mosaic}
\end{figure*}

In this section we describe the overall \hi structure of I~Zw~18.
We adopt the standard nomenclature introduced by \citet{Davidson1989}. The main body is
designated as I~Zw~18~A (Fig. \ref{fig:mosaic}, top-right) and is characterized by two
starburst regions: one to the North-West (NW) and one to the South-East (SE). The
light concentrations denoted by \citet{Davidson1989} with B, D and E are background galaxies.
The stellar complex to the NW is named I~Zw~18~C or C-component. We assume a distance
of 18.2 Mpc, as derived from the tip of the red giant branch \citep{Aloisi2007} and
confirmed by observations of Cepheids \citep{Fiorentino2010}.

We use data at three different resolutions (see table \ref{tab:cubes}) in order
to probe different spatial scales and \hi column densities.

\begin{figure*}[thp]
\centering
\includegraphics[width=18cm]{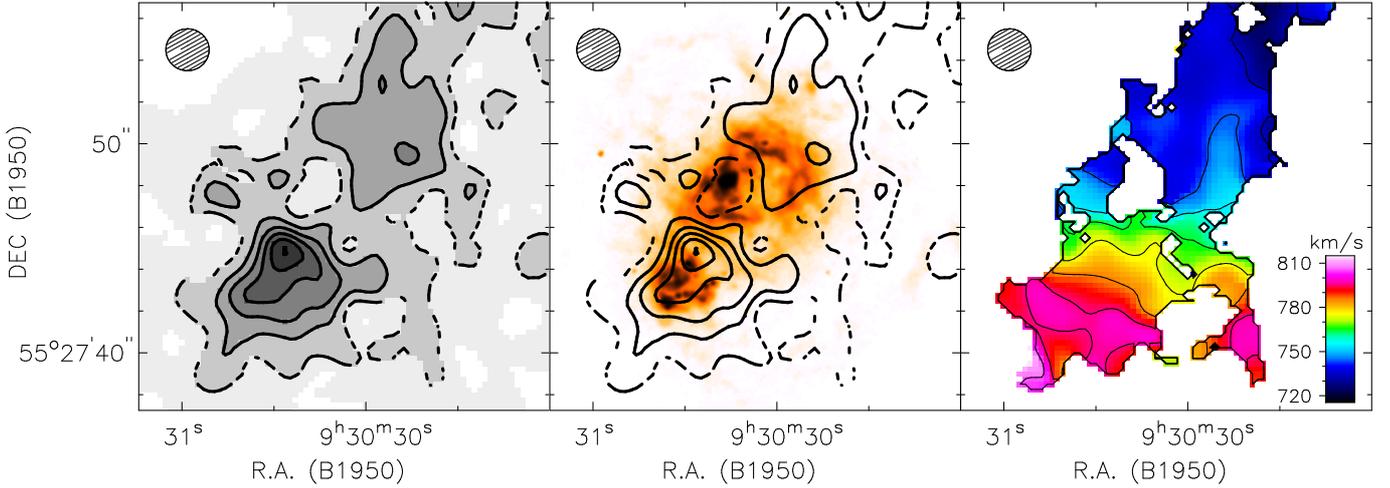}
\caption{\textbf{Left:} integrated \hi map at 2$''$ resolution. Countours are at 3 (dashed),
6, 9, 12, 15 $\times$ 10$^{21}$ atoms cm$^{-2}$. \textbf{Middle:} H$\alpha$ image (from \citealt{Cannon2002})
overlayed with the integrated \hi map at 2$''$ resolution. \textbf{Right:} velocity field at 2$''$ resolution.
Contours range from 722.4 to 805.6 \kms with steps of 10.4 km s$^{-1}$. The circle shows the beam size.}
\label{fig:s2mosaic}
\end{figure*}
\subsection{The low-resolution view \label{sec:s20}}

The low-resolution data (with FWHM = 20$''\sim$1.8 kpc and 3$\sigma$ column density
sensitivity N$_{\hi}(3\sigma)\sim4\times10^{18}$ atoms~cm$^{-2}$ per channel)
illustrate the large-scale overall structure of I~Zw~18.

Figure \ref{fig:s20} (top) shows the channel maps, overlayed with two isophotes of a B-band
image (from \citealt{GilDePaz2003}). The central \hi emission presents a velocity gradient
at a position angle P.A.$\simeq$140$^{\circ}$-150$^{\circ}$. At velocities from
$\sim$750 to $\sim$720 km s$^{-1}$ extended emission appears also to the South.

Figure \ref{fig:mosaic} (top-left) shows the integrated \hi map, overlayed on a V-band HST
image (from \citealt{Aloisi2007}). I~Zw~18~A is associated with a strong concentration of gas,
while diffuse emission extends beyond the optical galaxy, covering an angular size of
$\sim$3$'$.5 ($\sim$18.5 kpc). The \hi gas to the South of I~Zw~18~A displays
a tail-like morphology that extends over $\sim$2$'$.5 ($\sim$13.5~kpc).

Figure \ref{fig:mosaic} (middle-left) shows the velocity field. The main body (I~Zw~18~A) is
associated with the central velocity gradient. The southern ``tail'' does not seem to be kinematically
connected with the SE region of I~Zw~18~A, as the gas velocity changes abruptly from $\sim$790
km~s$^{-1}$ to $\sim$720 km~s$^{-1}$. Moreover, at the junction between I~Zw~18~A and the ``tail'',
the \hi line profiles are double peaked, suggesting that there are two distinct components,
possibly well-separated in space but projected to the same location on the sky.

Figure \ref{fig:mosaic} (bottom-left) shows a Position-Velocity (PV) diagram, obtained from
the 20$''$ datacube following the ``tail'' (the dashed line overlayed on the velocity field).
The central velocity gradient (associated with I~Zw~18~A) is very steep and there is a spatial
broadening towards the NW direction between $\sim$700 and $\sim$780 km~s$^{-1}$. The gas to the South
forms a coherent kinematical structure at velocities between $\sim$710 and $\sim$760 km~s$^{-1}$.
Strikingly, the broadened part of the PV-diagram and the ``tail'' are almost at the same velocities,
suggesting that they may be physically connected. This hypothesis will be investigated further
in Sect. \ref{sec:diffuse}.

\subsection{The intermediate-resolution view \label{sec:s5}}

The intermediate-resolution data (FWHM = $5''\sim440$~pc and N$_{\hi}(3\sigma)\sim4\times10^{19}$
atoms~cm$^{-2}$ per channel) illustrate the \hi emission associated with I~Zw~18~A and I~Zw~18~C,
and their possible connections.

Figure \ref{fig:s20} (bottom) shows the channel maps, overlayed with two isophotes of a
B-band image. The \hi emission to the NW is spatially resolved but still visible, whereas
the southern ``tail'' is completely resolved out, indicating the diffuse nature of this gas.
Between $\sim$770 and $\sim$740 km~s$^{-1}$ there are \hi clumps near the C-component.
Their association with I~Zw~18~C is very likely, because the \hi clumps are
at the same velocities as the H$\alpha$ emission
($V_{\rm{sys,\, \rm{H}\alpha}}[\rm{C}]=751 \pm 5$ km s$^{-1}$, \citealt{Dufour1996}).

Figure \ref{fig:mosaic} (top-right) shows the integrated \hi map, overlayed on a V-band
HST image. The main body is characterized by two \hi peaks, roughly corresponding to the
NW and SE starburst regions. With respect to I~Zw~18~A, the \hi is more extended to the
North-West, in the direction of I~Zw~18~C.

Figure \ref{fig:mosaic} (middle-right) shows the velocity field, while figure \ref{fig:mosaic}
(bottom-right) shows a PV-diagram obtained following the dashed line on the velocity field.
The steep velocity gradient is approximately along the two \hi peaks. The \hi emission to the
NW shows a shallow velocity gradient from I~Zw~18~A to I~Zw~18~C and seems to connect
the two stellar bodies. The presence of a ``gaseous bridge'' is supported also by H$\alpha$
observations: \citet{Dufour1990} detected diffuse H$\alpha$ emission connecting I~Zw~18~A and
I~Zw~18~C, while \citet{Dufour1996} measured H$\alpha$ velocities between the two bodies
that are very similar to those observed in \hi (their Fig. 3). Along I~Zw~18~A, instead,
the H$\alpha$ velocity gradient shows a ``wiggly'' behaviour that is not observed in \hiV.
This may be due to the presence of an H$\alpha$ superbubble (see \citealt{Martin1996}
and Sect. \ref{sec:Ha}). Also, there is \hi emission to the West of the main body, that
shows a velocity gradient and seems to have an H$\alpha$ counterpart (see Sect. \ref{sec:Ha}).

\subsection{The high-resolution view \label{sec:s2}}

\begin{figure*}[thp]
\centering
\includegraphics[width=17.5cm]{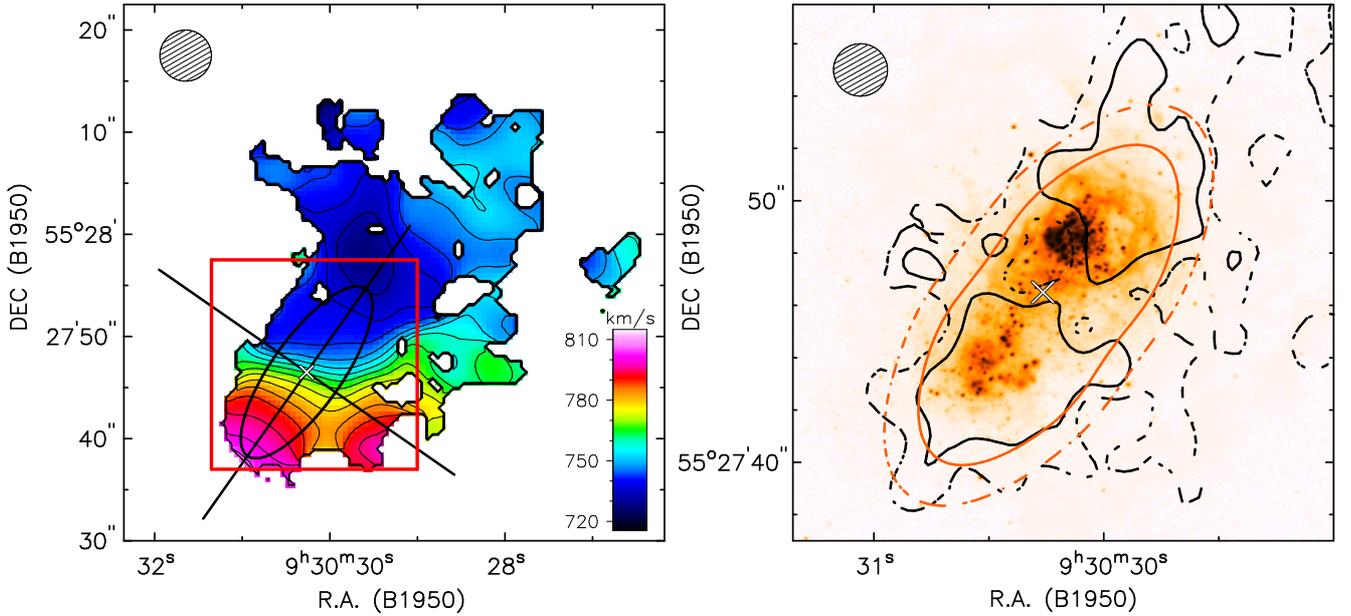}
\caption{\textbf{Left}: velocity field at $5''$ resolution. The ellipse shows
the modelled disk. The centre, the major and the minor axis are shown. The box
shows the area covered by the right panel. The circle shows the beam.
Countours are the same as in Fig. \ref{fig:mosaic}.
\textbf{Right}: HST image overlayed with the observed \hi map at 2$''$
(black) and the \hi map obtained from a 3D model with $i=70^{\circ}$ and
$z_{0} = 100$~pc (red). Countours are at 3 (dashed) and 6 (solid)
$\times$ 10$^{21}$ atoms cm$^{-2}$. The cross marks the centre. The circle shows the beam.}
\label{fig:sliceMap}
\end{figure*}
The high-resolution data (FWHM = $2''\sim180$ pc and N$_{\hi}(3\sigma)\sim2\times10^{20}$
atoms~cm$^{-2}$ per channel) show in detail
the \hi emission associated to the NW and SE starburst regions.

Figure \ref{fig:s2mosaic} shows the total \hi map in grayscale (left) and in contours
overlayed on a H$\alpha$ image (middle). The two \hi peaks are spatially resolved.
The \hi clump to the SE coincides with a complex of \hii regions and has a strong
peak, where the \hi column density reaches $\sim$100 M$_{\odot}$~pc$^{-2}$
($\sim 1.2 \times 10^{22}$ atoms cm$^{2}$).
In between the two clumps, an \hi hole is associated with a strong \hii region,
suggesting that the neutral gas has been consumed, ionized and/or blown out by
young stars. The \hi clump to the NW coincides with a H$\alpha$ shell, that
surrounds the bulk of the young stars (see Fig. 1 of \citealt{Cannon2002}).
This shell is probably connected with the high-velocity H$\alpha$ emission
detected by \citet{Dufour1996} and \citet{Martin1996} at $\pm$200 km~s$^{-1}$
with respect to the systemic velocity.

The velocity field at 2$''$ resolution is shown in Fig. \ref{fig:s2mosaic} (right).
This velocity field is very uncertain because of the clumpy \hi distribution, the asymmetric
line profiles and the low S/N-ratio of the data at this high angular resolution.
However, it shows a clear velocity gradient from the SE to the NW region, as already
observed at lower resolutions. The \hi depression is in the approaching NW side of the galaxy.

\section{Dynamics of I~Zw~18~A \label{sec:mainbody}}
In Sect. \ref{sec:cubes} we described the overall structure of I~Zw~18. Two main facts
need to be explained: i) the nature of the steep velocity gradient associated with I~Zw~18~A;
ii) the origin of the extended \hi emission to the South and to the North-West of the galaxy.
In this section we will focus on the dynamics of I~Zw~18~A, while in Sect. \ref{sec:diffuse}
we will study the large-scale gas emission.

\subsection{Kinematical models \label{sec:kin}}

There is some controversy in the literature regarding the velocity gradient of
I~Zw~18~A: \citet{Viallefond1987}, \citet{Petrosian1997} and \citet{vanZee1998}
analysed 2D velocity fields and interpreted the gradient as rotation, whereas
\citet{Skillman1993} and \citet{Dufour1996} obtained 1D long-slit spectroscopy
and argued that the gradient may result from the merger of two (or more) gaseous
clouds. The velocity gradient is along the optical major axis of the galaxy
(Fig. \ref{fig:sliceMap}) and the velocity field shows the typical pattern
due to rotation: this strongly suggests the presence of a rotating disk. Here
we present 3D kinematical models, which demonstrate that the \hi disk is
differentially rotating and has a global inflow/outflow motion.

\begin{figure*}[thp]
\centering
\includegraphics[width=17.5cm]{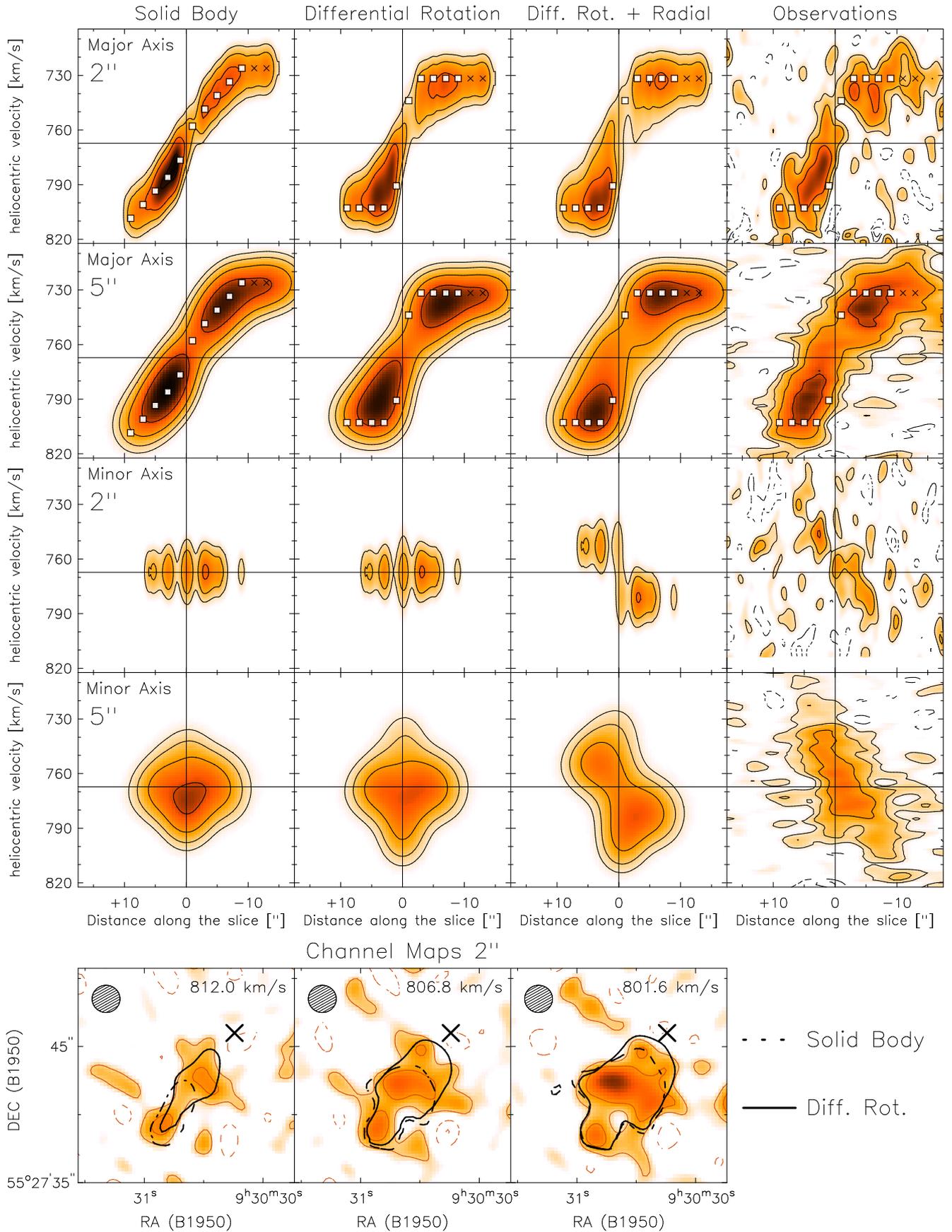}
\caption{Comparison between different 3D kinematical models and the observations.
\textbf{Top}: Position-Velocity diagrams at a resolution of 2$''$ and 5$''$. The slice
position are the major and minor axis, as indicated in Fig. \ref{fig:sliceMap} (left).
Contours are at -1.5 (dashed), 1.5, 3, 6, 12~$\times$~$\sigma$.
\textbf{Bottom}: Channels maps at 2$''$ resolution (red contours) overlayed with two models:
differential rotation (black-solid contours) and solid body rotation (black-dashed contours).
Contours are at 2$\sigma$. The cross marks the galaxy centre. The circle shows the beam size.
See text for details.}
\label{fig:models}
\end{figure*}
The disk is modelled by a set of gas rings, with fixed centre, systemic velocity,
position angle, inclination, surface density, thickness, velocity dispersion
and rotation velocity. The centre, the systemic velocity and the position angle
were estimated by eye using both optical and \hi data (see table \ref{tab:prop}).
The centre is between the NW and the SE starburst regions (see Fig. \ref{fig:sliceMap}, right).
For the radial distribution, we used the \hi surface density profile, derived from
the total \hi map at 2$''$ resolution by azimuthally averaging over ellipses (Fig.
\ref{fig:mass}, top). For the vertical distribution, we assumed an exponential
law $exp(-z/z_{0})$. We built a set of models with different values for the inclination
$i$, the scale height $z_{0}$ and the velocity dispersion $\sigma_{\hi}$, assuming that
each of these parameters is constant with radius.
The inclination and the scale height are constrained by the observed \hi map; their
values are slightly degenerated but do not strongly affect the final result; we assumed
$i=70^{\circ}$ and $z_{0} = 100$~pc (see Fig. \ref{fig:sliceMap}, right).
The mean velocity dispersion is constrained by the shape of different PV-diagrams and
values higher than $\sim$10 km~s$^{-1}$ are ruled out;  we assumed $\sigma_{\hi} = 7.5$
km~s$^{-1}$. A thickness of 100 pc and a \hi velocity dispersion of 7.5 km~s$^{-1}$ are
typical values for a \hi disk.

\begin{table}[thp]
\caption{Properties of I~Zw~18.}
\label{tab:prop}
\centering
\begin{tabular}{l c c}
\hline
\hline
Parameter                             & I~Zw~18~A                                       & I~Zw~18~C\\
\hline
$\alpha$ (B1950)                      & $09^{\rm{h}} 30^{\rm{m}} 30.^{\rm{s}}3 \pm 0.1$ & $09^{\rm{h}} 30^{\rm{m}} 27.^{\rm{s}}9 \pm 0.2$ \\
$\delta$ (B1950)                      & $55^{\circ} 27' 47'' \pm 1''$                   & $55^{\circ} 28' 06'' \pm 2''$ \\
$V_{\rm{sys}}$ (km s$^{-1}$)          & $767 \pm 4$                                     & $751 \pm 5$\\
Position Angle ($^{\circ}$)           & $145 \pm 5$                                     & -\\
Inclination Angle ($^{\circ}$)        & $70  \pm 4$                                     & -\\
$V_{\rm{rot}}$ (km s$^{-1}$)          & $38  \pm 4.4$                                   & -\\
$L_{\rm{B}}$ (10$^{7}$ L$_{\odot}$)   & 13.9                                            & 1.1\\
$L_{\rm{R}}$ (10$^{7}$ L$_{\odot}$)   & 5.8                                             & 0.4\\
$M_{\hi}$ (10$^{8}$ M$_{\odot}$)      & 1.0                                             & $\lesssim$1.2\\
$M_{\rm{dyn}}$ (10$^{8}$ M$_{\odot}$) & $3 \pm 1$                                       & -\\
\hline
\end{tabular}
\tablefoot{Optical luminosities were calculated using the apparent magnitudes
from \citet{Papaderos2002}, the distance from \citet{Aloisi2007} and the solar
absolute magnitudes from \citet{BinneyMerrifield1998}.
The \hi mass of I~Zw~18~C refers to the extended emission described in
Sect. \ref{sec:diffuse}, althought only part of this gas may be physically
associated with the C-component. The southern ``tail'' accounts for
$\sim0.5\times10^{8}$ M$_{\odot}$. }
\end{table}
The actual \hi distribution of I~Zw~18~A is clearly not axi-symmetric (see Fig. \ref{fig:s2mosaic}).
Thus, once the structural and geometrical parameters of the disk were fixed, we built
models with axi-symmetric kinematics but a clumpy \hi distribution,
i.e. the surface density varies with position as in the observed \hi map.
The procedure is as follows: we built a disk model with uniform density distribution
(fixing $z_{0}$, $\sigma_{\hi}$ and the rotation curve), projected it on the sky and then
renormalized the \hi line profiles at every spatial position in order to reproduce the \hi
density distribution observed at 2$''$ resolution. For the rotation curve, we tried two extreme cases:
solid body (slowly rising rotation curve) and differential (steeply rising and flat rotation curve).

Figure \ref{fig:models} (top) shows PV-diagrams obtained from different models and from
the observations at a resolution of 2$''$ and 5$''$. The slice
positions are the major and minor axis, as indicated in Fig. \ref{fig:sliceMap} (left).
The velocity gradient along the major axis is grossly reproduced by all the models. This
demonstrates that: i) a rotating disk is a good representation of the data; ii) the asymmetry
between the NW and the SE region is mostly caused by the clumpy \hi distribution. Moreover,
it is possible to discriminate between solid body and differential rotation. The observed
PVD along the major axis shows \hi emission close to the galaxy centre ($R < 5''$) at high
rotational velocities ($\sim$810 and $\sim$730 km s$^{-1}$). The solid body model does not
reproduce such emission, as the gas in the inner radii is mostly concentrated near the
systemic velocity. The differentially rotating model, instead, correctly reproduces
the high-velocity gas. This is clearly illustrated by the channel maps at 2$''$ resolution
at receding velocities (Fig. \ref{fig:models}, bottom): the solid body model (dashed line) is not
extended enough towards the galaxy centre, whereas the differentially rotating model
(solid line) gives a good match with the observations. The approaching NW side of the galaxy
is not reproduced as well as the receding SE side. The NW starburst region is more active
than the SE one \citep{ContrerasRamos2011} and this may affect its local kinematics. Nonetheless,
also on this side, a differentially rotating disk is preferable to a solid body one.

A simple rotating disk (independently from the assumed rotation curve) cannot reproduce
the observed PV-diagrams along the minor axis (Fig. \ref{fig:models}, top), because of the
kinematic asymmetry and the presence of \hi emission at velocities forbidden by circular motions.
This effect is visible also in the velocity field at 5$''$ resolution (Fig. \ref{fig:sliceMap},
left): the kinematic minor axis, defined by the contours close to the systemic velocity,
is not orthogonal to the kinematic major axis. Usually, this is attributed to radial motions
(e.g. \citealt{Fraternali2002}). Alternatively, the non-orthogonality between the minor and
major axes may indicate the presence of a bar-like or oval distortion (e.g. \citealt{Bosma1978}).
The optical images of I~Zw~18, however, do not indicate the presence of such strong distortions.
We improved the differentially rotating model by adding a global radial motion of 15 km~s$^{-1}$.
The resulting model reproduces the \hi emission at forbidden velocities. The model, however,
cannot reproduce all the details present in the observed PV-diagram. The non-circular motions
are not uniform across the disk and small variations (of the order of 3-4 km~s$^{-1}$)
could account for the observed discrepancies. It is not possible to discriminate
between inflow and outflow, as it is not known which side of the disk is the near one.
Vertical motions with roughly the same speed as the radial ones may also be present.
Non-circular motions in excess of 20 km~s$^{-1}$ are ruled out, confirming that
the disk kinematics is dominated by rotation.

The rotation curve used to build the best model was not derived from a standard
tilted-ring fit of the velocity field \citep{Begeman1987}, but it was derived with
a trial and error approach building a series of 3D models. The uncertainties on
the rotation velocities are difficult to quantify. We made a conservative estimate
of the uncertainties assigning an error equal to $\Delta V /2.35 = 4.4$ km~s$^{-1}$,
where $\Delta V$ is the velocity resolution of the 2$''$ datacube. The first point
of the rotation curve is the most uncertain, as the velocity dispersion in the inner
ring may be higher than the mean value of 7.5 km~s$^{-1}$. For example, if $\sigma_{\hi}$
is assumed to be 10 km~s$^{-1}$ higher than the mean value, the rotation velocity
would decrease of $\sim$5 km~s$^{-1}$. We calculated the asymmetric drift correction
following \citet{Meurer1996}, but it turns out to be smaller than the errors.

\subsection{Mass models \label{sec:dyn}}

In Sect. \ref{sec:kin} we showed that I~Zw~18~A has a rotating \hi disk.
The rotation curve is uncertain, but it shows an inner steep rise and an
outer flat part. This indicates the presence of a strong central concentration
of mass, that may be luminous or dark. Using this rotation curve,
we built mass models to estimate the relative contributions of luminous
and dark matter to the gravitational potential. We followed \citet{Begeman1987}.

\begin{figure}[htbp]
\centering
\includegraphics[width=12cm, angle=-90]{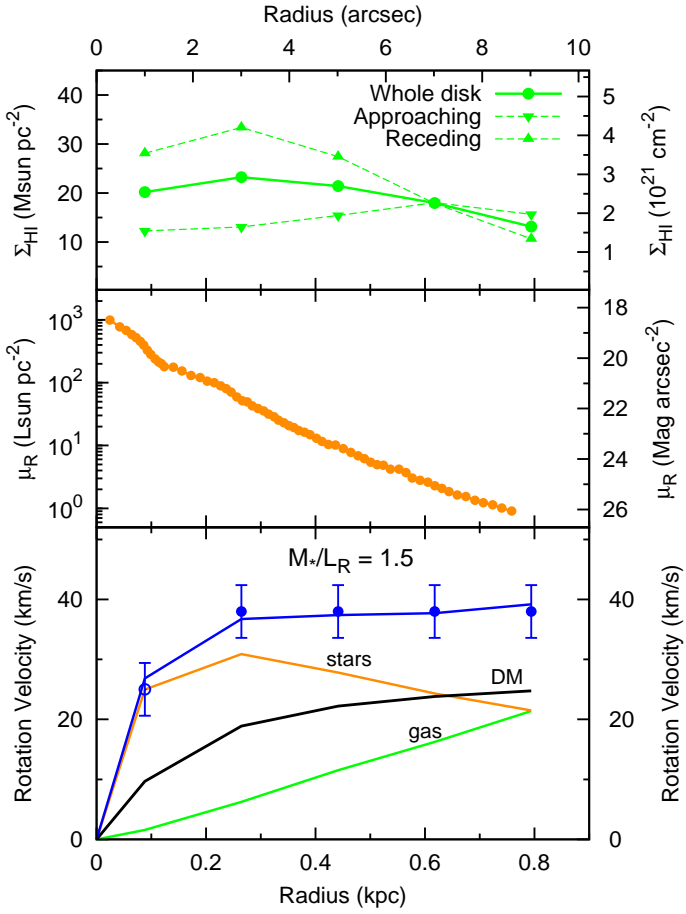}
\caption{\textbf{Top}: \hi surface density profile (inclination corrected),
derived by ellipse averaging over the entire disk (dots) and over the approaching
and receding sides separately (down-triangles and up-triangles, respectively).
\textbf{Middle:} R-band surface brightness profile (from \citealt{Papaderos2002}).
\textbf{Bottom:} ``maximum disk'' decomposition of the rotation curve. Dots and
circles show the observed curve. Lines show the contributions due to gas, stars
and dark matter and the resulting rotation curve.}
\label{fig:mass}
\end{figure}

The contribution of the gaseous disk was computed using the surface density
profile derived from the total \hi map at 2$''$ resolution (Fig. \ref{fig:mass}, top),
multiplied by a factor 1.4 to take into account the presence of Helium.
The possible gravitational effect of the \hi components \textit{outside} the disk
(i.e. the ``tail'' to the South and the extensions to the North-West and to the West)
was not taken into account. Molecular gas was not explicitly considered in the
mass model because its amount is very uncertain (\citealt{Leroy2007}).
However, if molecules are distributed as the stars, their contribution
is reflected in an increase of the stellar mass-to-light ratio (M$_{*}$/L).
Consistently with the models in Sect. \ref{sec:kin}, we assumed an exponential
vertical distribution with $z_{0}=100$~pc.

The contribution of the stars was computed using the R-band surface brightness
profile from \citet{Papaderos2002} (Fig. \ref{fig:mass}, middle), that is derived
from an HST image after the subtraction of the nebular emission (dominated by
the H$\alpha$ line). The color profiles from \citet{Papaderos2002} (their Fig. 11)
show that, after the subtraction of the nebular emission, the color of I~Zw~18~A is
almost constant with radius. Thus, it makes sense to use a constant value of
M$_{*}$/L$_{\rm{R}}$. We assumed a stellar disk with a vertical density distribution
given by $\rho(z)=\rm{sech}^{2}(z/z_{0})$ \citep{vanDerKruit1981}, with $z_{0} = 100$ pc.

For the dark matter distribution, we assumed a pseudo-isothermal halo described
by equation:
\begin{equation} \label{eq:ISO}
 \rho_{\rm{ISO}}(r) = \frac{\rho_{\rm{0}}}{1 + (r/r_{\rm{c}})^{2}},
\end{equation}
where $\rho_{\rm{0}}$ is the central density and $r_{\rm{c}}$ is the core radius.
$\rho_{\rm{0}}$ and $r_{\rm{c}}$ are free parameters of the mass models.

Figure \ref{fig:mass} (bottom) shows the ``maximum disk'' decomposition of the
rotation curve. The resulting stellar mass-to-light ratio is M$_{*}$/L$_{\rm{R}}$~$\sim$~1.5.
In the maximum disk hypothesis, the baryons dominate the gravitational
potential in the inner regions, while the dark matter halo dominates in the
outer parts. The parameters of the halo are uncertain because the sampling
of the rotation curve is poor. The halo shown in Fig. \ref{fig:mass} (bottom) has:
$\rho_{0}= 833 \times 10^{-3}$~M$_{\odot}$~pc$^{-3}$ and $r_{\rm{c}}=0.13$~kpc.

A M$_{*}$/L$_{R}$~$\sim$1.5 implies a stellar mass of
$\sim$9$\times$10$^{7}$~M$_{\odot}$. According to \citet{Aloisi1999},
the starburst started $\sim$20~Myr ago with a star-formation rate
(SFR) of 6$\times$10$^{-2}$ M$_{\odot}$~yr$^{-1}$, giving a mass in
\textit{young stars} of $\sim$10$^{6}$~M$_{\odot}$\footnote{
\citet{Aloisi1999} assumed a distance of 10~Mpc. Considering the new
value of 18.2~Mpc, the starburst parameters slightly change, but the mass
in young stars remains almost the same (F. Annibali, priv. comm.).}.
Thus, the newly formed stars and the concentration of \hi cannot
explain the steep rise of the rotation curve, implying that the
mass concentration is due to either \textit{old} stars, or molecules,
or dark matter. Old stars have been detected \citep{Aloisi2007}
and their total mass can be constrained by deriving the galaxy Star-Formation
History from Color-Magnitude diagrams. The maximum-disk value requires
a mean SFR of $\sim$7$\times$10$^{-3}$ M$_{\odot}$~yr$^{-1}$ over
the last 13 Gyr, that cannot be ruled out. Regarding molecules, the
upper limit for the H$_{2}$ mass within $\sim$400~pc is $\sim 7 \times 10^{5}$ M$_{\odot}$
\citep{Leroy2007}\footnote{We rescaled the original values to a distance of 18.2 Mpc.},
using a Galactic CO-to-H$_{2}$ conversion factor (X$_{\rm{CO}}$).
However, \citet{Leroy2007} argued that, in I~Zw~18, X$_{\rm{CO}}$
may be 10$^{-2}$ times the Galactic value. The same result is found by
extrapolating the relation between X$_{\rm{CO}}$ and metallicity
by \citet{Boselli2002} down to the metallicity of I~Zw~18.
Thus, the H$_{2}$ mass within $\sim$400 pc may be dynamically
important and as high as $\sim$7$\times$10$^{7}$~M$_{\odot}$.

We also considered MOdified Newtonian Dynamics (MOND) (\citealt{Milgrom1983},
\citealt{Sanders2002}) to describe the rotation curve. We fixed $a_{0} = 1.21 \times 10^{-8}$
cm~s$^{-2}$ \citep{Begeman1991} and the distance $D=18.2$ Mpc \citep{Aloisi2007},
thus the only free parameter is M$_{*}$/L$_{\rm{R}}$. MOND provides acceptable
fits and gives M$_{*}$/L$_{\rm{R}} = 1.5$ using the ``standard'' interpolation
function \citep{Milgrom1983} and M$_{*}$/L$_{\rm{R}} = 1$ using the ``simple''
one \citep{Famaey2005}.

Following \citet{McGaugh2011}, we check the position of I~Zw~18~A on the baryonic
Tully-Fisher relation. The galaxy follows the correlation within the observed scatter.

\section{The extended emission \label{sec:diffuse}}

In this section we study the extended \hi emission. This may provide some clues
to the mechanism that triggered the starburst. Also, we compare the
large-scale \hi and H$\alpha$ emission to investigate the possible presence of outflows.

\subsection{The C-component and the \hi tail \label{sec:tail}}

\begin{figure}[tbp]
\centering
\includegraphics[width=8cm]{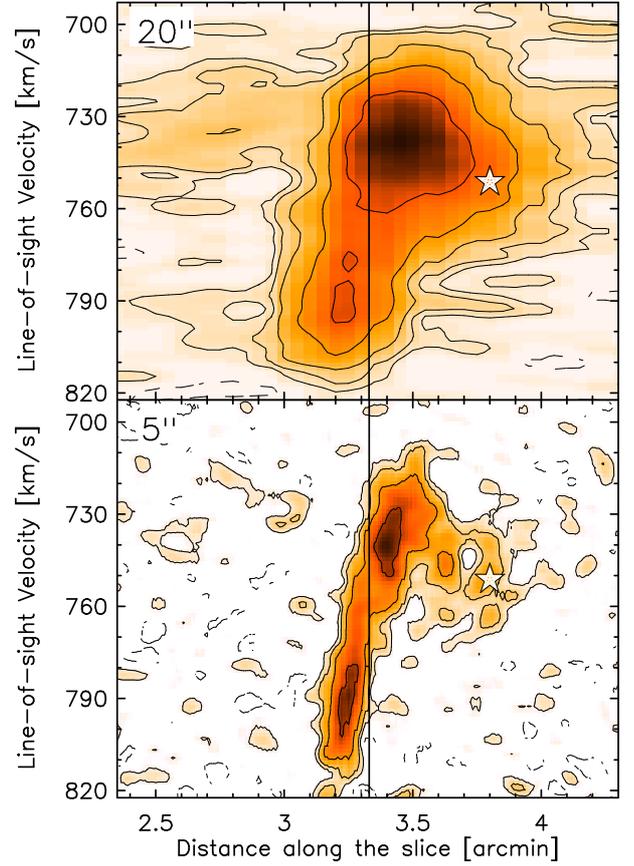}
\caption{Position-Velocity diagrams at a resolution of 20$''$ (top) and 5$''$
(bottom). They are derived following the path shown in Fig. \ref{fig:mosaic}.
Countours are at -1.5 (dashed), 1.5, 3, 6, 12, 24 $\sigma$. The star shows
the H$\alpha$ velocity and the spatial position of I~Zw~18~C.}
\label{fig:PVD}
\end{figure}
\begin{figure*}[thp]
\centering
\includegraphics[width=18cm]{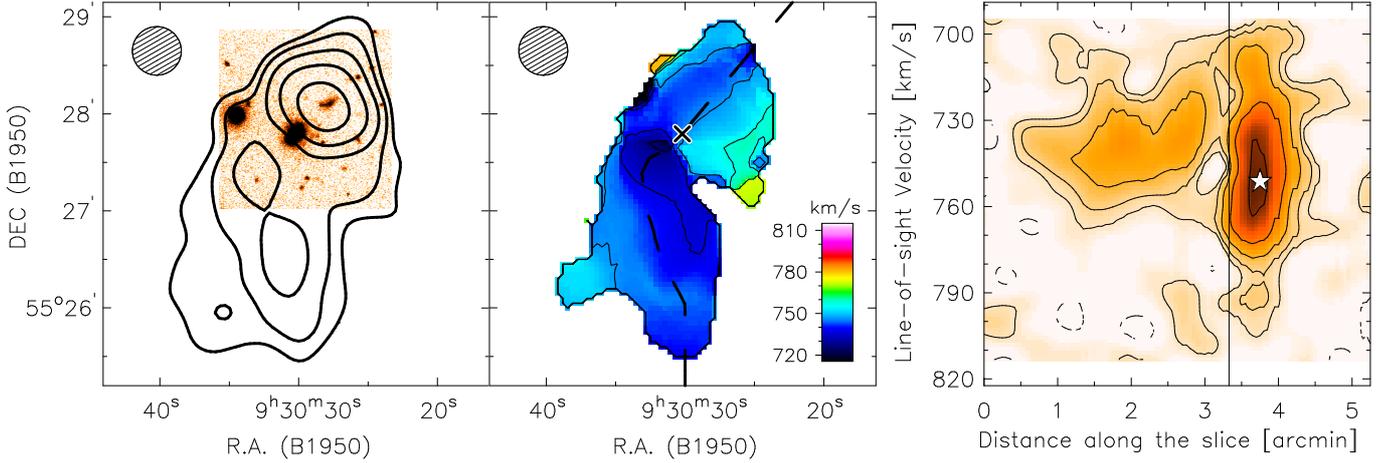}
\caption{The \hi emission at 30$''$ resolution, after the subtraction
of the main body.
\textbf{Left}: B-band image overlayed with the total \hi map. Countours are at
2.4, 4.8, 9.6, 19.2 $\times$ 10$^{19}$ atoms cm$^{-2}$. \textbf{Middle}:
Velocity field. Contours range from 735.8 to 767 \kms with steps of 10.4 km s$^{-1}$.
The cross shows the position of I~Zw~18~A. The dashed line shows the path followed
to obtain the position-velocity diagram. \textbf{Right}: Position-Velocity diagram.
Countours are at -1.5 (dashed), 1.5, 3, 6, 12 $\sigma$, where $\sigma=0.3$ mJy/beam.
The vertical line corresponds to the cross in the velocity field. The star shows
the H$\alpha$ velocity and the spatial position of I~Zw~18~C.}
\label{fig:diffuse}
\end{figure*}
In Sect. \ref{sec:s20} we reported two puzzling results (see Fig. \ref{fig:mosaic}, left):
i)~an \hi ``tail'' at velocities between 710-760 km~s$^{-1}$ extending to the South of
I~Zw~18~A and kinematically disconnected from the South-East side of the central rotating
disk; ii)~a broadening of the \hi emission in the PV-diagram almost at the same velocities
(700-780 km~s$^{-1}$) to the North-West (in the direction of I~Zw~18~C). This is clearly
shown in Fig. \ref{fig:PVD}, where PV-diagrams at 20$''$ and 5$''$ resolution (with
different column density sensitivities) are plotted on the same scale.

In order to study these components and their possible connection in more detail,
we subtracted the compact \hi disk of I~Zw~18~A from the surrounding extended
\hi emission. We used the high-resolution data to define the emission from the disk
and we subtracted this emission from the low-resolution datacube\footnote{
Technically, we built a mask containing only the \hi signal from the disk and
we cleaned the datacube at $1.5''\times1.4''$ resolution down to 1$\sigma$,
using the mask to define the search areas. We restored the clean-components
on a blank cube, using a Gaussian beam of 20$''$. The resulting ``clean-components''
cube contains only the emission from the disk, but at the desired resolution
of 20$''$. Finally, the ``clean-components'' cube was
subtractred channel by channel from the 20$''$ datacube.}.
Subsequently, we smoothed the residual datacube to 30$''$ and to 10.4 km s$^{-1}$
and used it to obtain:
i) a total \hi map by summing the channels in the velocity range
$\sim$700-780 km s$^{-1}$; ii) a velocity field by estimating an
intensity-weighted mean velocity; iii) a PV-diagram by following the tail
(dashed line in Fig. \ref{fig:diffuse}).

Figure \ref{fig:diffuse} shows the results of the subtraction. Interestingly,
extended \hi emission is centered on I~Zw~18~C (left panel) and forms
a coherent kinematical structure with velocities between $\sim$700 and 800
km s$^{-1}$ (right panel). The physical association of the C-component
with this surrounding \hi emission is likely because the H$\alpha$ systemic
velocity of I~Zw~18~C is $\sim$751$\pm$5 km~s$^{-1}$ \citep{Dufour1996},
(see star in Fig. \ref{fig:diffuse}, right).

The southern \hi tail seems to be connected in space and in velocity with
the \hi structure around I~Zw~18~C, as is shown by the velocity field
(Fig. \ref{fig:diffuse}, middle). The connection may be behind or in front
of I~Zw~18~A. Possible interpretations of the extended \hi emission will
be discussed in Sect. \ref{sec:inter}.

\begin{figure}[htbp]
\centering
\includegraphics[width=7cm, angle=-90]{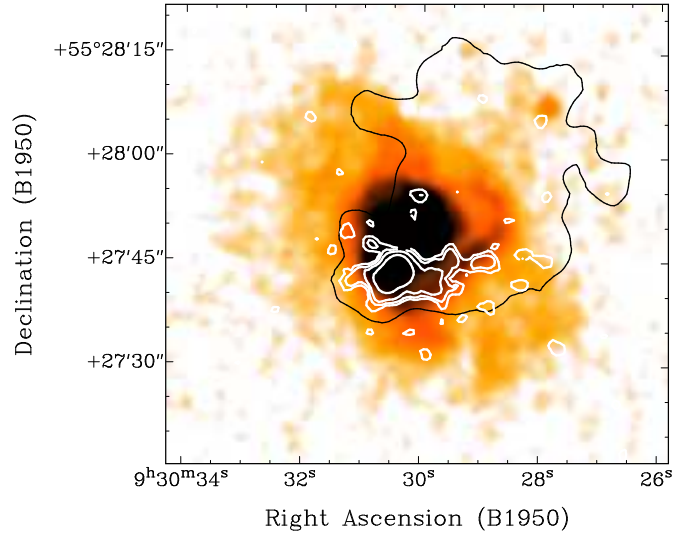}
\caption{H$\alpha$ image (\citealt{GilDePaz2003}) overlayed with the \hi emission.
White contours show the \hi emission at 2.$''8\times3.''$3 resolution, integrated in the
velocity range 800-770 km s$^{-1}$, and correspond to 5, 10, 20, 40 $\times$ 10$^{20}$
atoms cm$^{-2}$. The black line shows the pseudo-3$\sigma$ contour of the total \hi map
at 5$''$ res.}
\label{fig:Ha}
\end{figure}
\subsection{The connection between \hi and $\rm{H}\alpha$ emission \label{sec:Ha}}

In Sect. \ref{sec:s2} we pointed out the relative \hi and H$\alpha$
distributions in the inner regions of I~Zw~18~A. Here we compare the
distribution and kinematics of neutral and ionized gas on larger scales.

Figure \ref{fig:Ha} shows an H$\alpha$ image (from \citealt{GilDePaz2003})
overlayed with the \hi emission (black and white contours).
The H$\alpha$ emission extends well beyond the stellar body and is almost
perpendicular to the \hi disk, suggesting the presence of an outflow.
This interpretation is in agreement with the H$\alpha$ long-slit spectroscopy
by \citet{Martin1996}, who detected a bipolar superbubble expanding
with velocities of $\pm$60 km~s$^{-1}$ out to $\sim$2 kpc from I~Zw~18~A.
The \hi to the North-East of the main body (see black contour) seems to
border the H$\alpha$ emission and is almost at the same velocities, suggesting
that part of the diffuse \hi may be associated with the outflow.

The H$\alpha$ emission presents also a prominent arc to the West of I~Zw~18~A.
This feature was first identified by \citet{Dufour1990} and interpreted as a
radiation-bound ionization front driven into the ISM. \citet{Petrosian1997}, instead,
argued that the H$\alpha$ arc contains also stellar emission and suggested
that it is a structure with stars able to ionize the gas ``in situ''.
The \hi to the West of I~Zw~18~A is associated with the H$\alpha$ arc. To show this,
we summed the channel maps at 2.$''8\times3.''$3 resolution in the velocity range
800-770 km~s$^{-1}$ (see Fig. \ref{fig:Ha}, white contours). Moreover, the H$\alpha$
velocity field of \citet{Petrosian1997} shows a gradient along the arc similar to
the one observed in Fig. \ref{fig:mosaic} (middle-right), confirming the
physical association between \hi and H$\alpha$. The hypothesis of a radiation-bound
ionization front is difficult to reconcile with the presence along the H$\alpha$ arc
of high-density ($\sim$10$^{21}$ atoms cm$^{-2}$) neutral gas.

\section{Discussion}

\subsection{Observational evidence \& interpretation \label{sec:inter}}

This \hi study of I~Zw~18 has shown that:
\begin{itemize}
 \item I~Zw~18~A has a compact rotating disk with very high \hi densities.
       The rotation curve is flat with a steep inner rise,
       indicating the presence of a strong concentration of mass.
       Also, a global inflow/outflow motion is present;
 \item I~Zw~18~C is located in the direction of the major axis of I~Zw~18~A
       and is almost at the same velocities as its approaching side.
       Gas emission with a smooth velocity gradient connects the two stellar bodies.
       I~Zw~18~C appears to be at the centre of a diffuse \hi structure;
 \item a \hi tail extends to the South of I~Zw~18~A out to $\sim$13.5~kpc.
       The tail has a coherent kinematical structure and seems to be connected
       with the \hi emission to the North-West.
\end{itemize}
Studies of the resolved stellar populations \citep{Aloisi2007, ContrerasRamos2011}
have shown that: i) the two starburst regions in I~Zw~18~A (NW and SE) are embedded
in a common envelope of old stars with ages $>$1~Gyr; ii) I~Zw~18~A and I~Zw~18~C
are two completely separate stellar bodies and there are no stars between them;
iii) also I~Zw~18~C contains both old ($>$1 Gyr) and young ($\sim$10~Myr) stars,
but its current star formation rate (SFR) is lower than that of I~Zw~18~A.

For the interpretation, we consider first the hypothesis of an interaction/merger
of two (or more) gas-rich dwarfs. It is well-known that interactions/mergers can
produce tidal tails (e.g. \citealt{Toomre1972}). Also, numerical simulations
(e.g. \citealt{HibbardMihos1995}) suggest that mergers can lead to gas inflows,
produce strong gas concentrations and trigger intense star-formation. Thus,
an interaction/merger may provide an explanation for:
i) the concentration of \hiV, ii) the on-going starburst, and iii)
the southern \hi tail. I~Zw~18~C may be either a ``relic'' of the interaction or a dwarf
galaxy that is interacting/merging with I~Zw~18~A. The two objects are at a projected
distance of $\sim$2.2 kpc, the difference between their systemic velocities
is $\sim$12 km~s$^{-1}$, and they are connected by \hi emission with a smooth
velocity gradient. The ratio between the R-band luminosities of I~Zw~18~A and
I~Zw~18~C is $\sim$14, thus this would be classified as a minor merger. The
merger hypothesis may also explain the extremely low metallicity, as
discussed in the following.

\citet{Bekki2008} argued that BCDs with low nebular metallicity are the results
of mergers between gas-rich dwarfs with extended \hi disks. According to his
simulations, the central starburst is fuelled with metal-poor gas transferred
from the outer regions of the extended disks, where the star formation and the
chemical enrichment were not efficient due to the low \hi densities.
I~Zw~18 is consistent with this picture, as the C-component is surrounded
by an extended \hi structure, that does not have a stellar counterpart and,
likely, has not been efficiently enriched by SN explosions. Therefore, this
\hi structure may provide ``fresh'' unprocessed gas into the starburst regions
of I~Zw~18~A. A similar mechanism of metal dilution has been proposed also by
\citet{Ekta2010b} to explain why I~Zw~18 and the other extremely metal-deficient
BCDs are outliers of the mass-metallicity relation.

The presence of the extended \hi emission, including the tail, has been considered
above as supporting evidence for the merger, but could it be instead the result of
a blowout from the starburst? Indeed, H$\alpha$ observations of I~Zw~18 suggest the
presence of such an outflow (Sect. \ref{sec:Ha}). Also, numerical simulations
predict that starbursting dwarfs undergo massive ouflows
because they have a shallow gravitational potential (e.g. \citealt{MacLow1999}).
The rate of the outflowing gas $dM_{\rm{out}}/dt$ can be roughly estimated as:
\begin{equation}
\frac{dM_{\rm{out}}}{dt} = \frac{2 \times \varepsilon \times SNR \, \times \overline{E_{\rm{SN}}}} {V_{\rm{esc}}^{2}}
\end{equation}
where $SNR$ is the rate of supernovae (SN), $\overline{E_{\rm{SN}}}$ is the mean
energy of a SN, $V_{\rm{esc}}$ is the escape velocity and $\varepsilon$
is the efficiency of the SN-feedback. Thus, $M_{\rm{out}} = dM_{\rm{out}}
/dt \times \Delta T$, where $\Delta T$ is the duration of the starburst.
Assuming $\overline{E_{\rm{SN}}}= 1.2 \times 10^{51}$ erg, $V_{\rm{esc}}=
\sqrt{2} \times V_{\rm{rot}}$, $\varepsilon = 0.15$, $\Delta T = 20$ Myr
\citep{Aloisi1999} and $SNR = 0.01 \times SFR$ with $SFR = 0.06$ M$_{\odot}$~yr$^{-1}$
\citep{Aloisi1999}, we get $M_{\rm{out}}\sim 5-8 \times 10^{7}$~M$_{\odot}$.
The total mass of the extended gas is $\sim1.6\times10^{8}$ M$_{\odot}$
(corrected for the presence of He). Thus, a massive outflow may explain
\textit{all} the diffuse gas if only slightly higher values of $SFR$, $\Delta T$
and $\varepsilon$ are assumed. Since the extended \hi emission is entirely
at approaching velocities, any outflow should be highly asymmetric and confined.

Finally, there is also the hypothesis of a ``fragmenting \hi cloud in the early
stages of galaxy evolution'', which was suggested by \citet{vanZee1998}. This
picture can explain the extremely low metallicity of I~Zw~18, but is in contrast
with the results of \citet{Aloisi2007}, who concluded that I~Zw~18 has old stars
and is not a young galaxy in formation. Alternatively, I~Zw~18~A and I~Zw~18~C may
be old stellar systems that are accreting cold gas from the inter-galactic medium
and are now forming new stars. This may be in line with some simulations of
dwarf galaxy formation (e.g. \citealt{Keres2005, Dekel2006}).

\subsection{Comparison with other dwarf galaxies}

The evolution of BCDs is still an open issue. In particular, it is not clear what
objects can be identified as their progenitors and descendants (\citealt{Papaderos1996,
vanZee2001}). It is useful, therefore, to compare their properties with those of
other types of dwarf galaxies.

\begin{figure}[thbp]
\centering
\includegraphics[width=14cm, angle=-90]{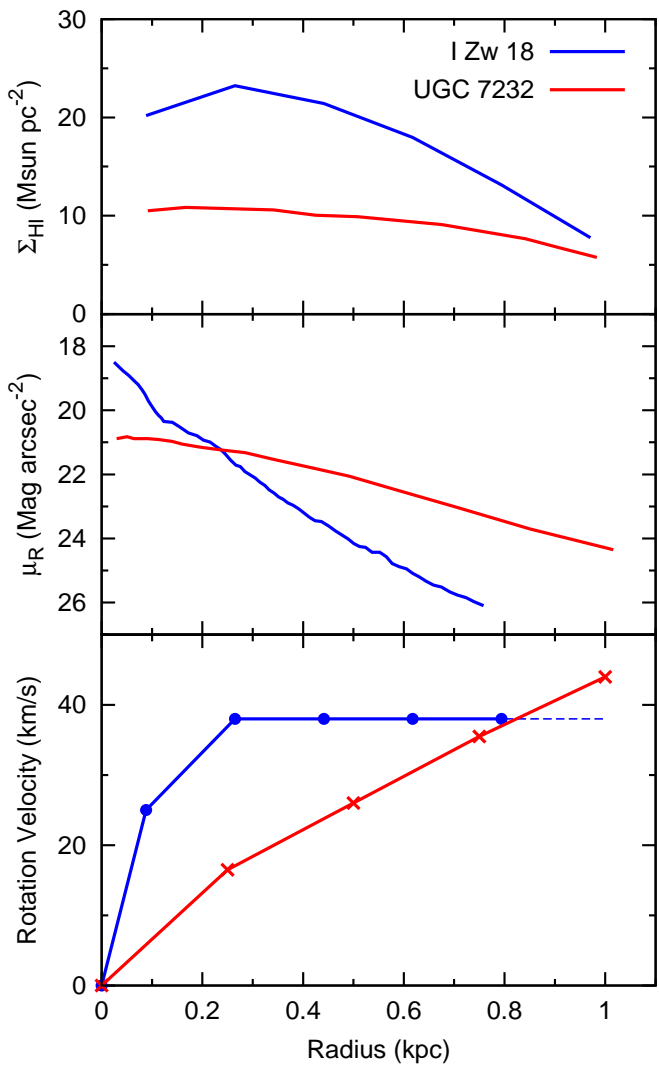}
\caption{Comparison between I~Zw~18~A (blue line) and a typical dwarf irregular
UGC~7232 (red line), selected from the sample of \citet{Swaters2009}. \textbf{Top:}
\hi surface density profile. \textbf{Middle:} R-band surface brightness profile.
\textbf{Bottom}: \hi rotation curve.}
\label{fig:comp}
\end{figure}
In Fig. \ref{fig:comp} we compare I~Zw~18~A with a typical dwarf irregular (UGC~7232),
taken from the sample of \citet{Swaters2009}. The \hi observations for these two
objects have almost the same linear resolution ($\sim$200~pc), making it possible
to compare \hi surface densities and velocity gradients. The two galaxies have
approximately the same \hi size and the same rotation velocity at the last measured
point, thus they have roughly the same dynamical mass. However, their structural properties
are very different:
\begin{enumerate}
 \item the azimuthally averaged \hi surface densities of I~Zw~18~A are a factor $\sim$2
       higher than those of UGC~7232 (Fig \ref{fig:comp}, top). Also, the \hi distribution
       of I~Zw~18~A is clumpy with \hi column densities as high as $\sim$50-100
       M$_{\odot}$~pc$^{-2}$ (Fig. \ref{fig:s2mosaic});
 \item the stellar component of I~Zw~18~A is much more compact than that of UGC~7232
       (Fig. \ref{fig:comp}, middle);
 \item the rotation curves have a completely different shape (Fig. \ref{fig:comp}, bottom).
       UGC~7232 has a slowly rising rotation curve, indicating a smooth mass distribution.
       I~Zw~18~A, instead, has a flat rotation curve with an inner steep rise,
       indicating the presence of a strong concentration of mass.
\end{enumerate}
In Section \ref{sec:dyn} we showed that the central mass concentration cannot be explained
by the newly formed stars and/or by the concentration of HI, but it can be identified with
the old stars and/or dark matter. As to the molecules, their amount is highly uncertain and
it is not clear whether they are dynamically important or not. The now uncovered concentration
of mass is unique amoung dwarf irregular galaxies and must be tightly linked with the
starburst. Furthermore, this result sheds new light on the question of the evolution of BCDs
and of their descendants. It is clear that, unless a significant re-distribution
of mass takes place, a steeply rising rotation curve, as found for I Zw 18, would
be the distinctive signature that makes the descendants recognizable.

Regarding other BCDs, it is known that their underlying stellar component, which
is formed by \textit{old stars}, is generally more compact than common dEs and dIrrs
(e.g. \citealt{Papaderos1996, GilDePaz2005}). In particular, the old stellar component
of the majority of BCDs has a typical central surface brightness $\mu_{0}\sim21$
mag arcsec$^{-2}$ in the B-band (e.g. \citealt{GilDePaz2005}), similar to high
surface brightness (HSB) disk galaxies (e.g. \citealt{vanderkruit2011}). If the
distribution of mass is strongly coupled to the distribution of light
(\citealt{Sancisi2004}), we expect that BCDs show a dynamical behaviour similar
to that of HSB spiral galaxies, i.e. steeply rising rotation curves that can be
described under the maximum disk hypothesis. This seems to be the case for I~Zw~18~A.
In all these respects, I~Zw~18~A resembles a ``miniature'' HSB disk galaxy. There
are already indications that BCDs may have ``steeper rotation curves than similar
luminosity, low surface brightness dwarf galaxies'' (\citealt{vanZee2001}),
but a detailed dynamical study is needed to trace reliable rotation curves
and determine the relative contributions of gas, stars and dark matter to
the gravitational potential.

\section{Conclusions}

We analysed \hi observations of the blue compact dwarf galaxy I~Zw~18.
Our main results can be summarized as follows:
\begin{itemize}
 \item the \hi associated with the starburst region (I~Zw~18~A) is in a compact rotating
       disk. The \hi column densities are very high, up to $\sim$50-100~M$_{\odot}$~pc$^{-2}$
       ($\sim0.6-1.2\times10^{22}$ atoms cm$^{-2}$);
 \item the disk has a flat rotation curve with an inner steep rise. This indicates
       the presence of a strong concentration of mass, that may be luminous or dark.
       Baryons may dominate the gravitational potential in the inner regions;
 \item the disk has a radial inflow/outflow motion of $\sim$15 km~s$^{-1}$;
 \item the stellar concentration to the North-West (I~Zw~18~C) is surrounded
       by extended \hi emission, smoothly connected with I~Zw~18~A;
 \item a \hi tail extends to the South of I~Zw~18~A out to $\sim$13.5 kpc.
       It shows a coherent kinematical structure and seems to be connected with
       the \hi emission to the North-West.
\end{itemize}
I~Zw~18~A appears structurally different from a typical dIrr in terms of \hi distribution,
stellar distribution and dynamics. In particular, it has a strong central concentration
of mass. It may be considered as ``miniature'' HSB disk galaxy. The \hi concentration and
the dynamical properties must be tightly linked with the starburst. They are also crucial
to address the question of the progenitors/descendants of BCDs.

Regarding the mechanism that triggered the starburst, an interaction/merger between
gas-rich dwarf galaxies seems to be the most likely hypothesis.

\begin{acknowledgements}
We thank F. Annibali, G. Fiorentino and M. Tosi for helpful discussions
about the stellar populations of BCDs and for providing the HST
image of I~Zw~18. We are grateful to J. van Gorkom for stimulating discussions.
Also, we thank J. Cannon for kindly making his HST images available to us.
\end{acknowledgements}
\bibliographystyle{aa}
\bibliography{17867bib.bib}
\end{document}